\documentclass{aa}
\usepackage{graphics}

\newcommand{\bea}{\begin{eqnarray}}
\newcommand{\eea}{\end{eqnarray}}
\newcommand{\be}{\begin{equation}}
\newcommand{\ee}{\end{equation}}
\begin{document}
\thesaurus{01    
        (11.02.2 AO 0235+164; 11.16.1; 02.18.5)}
\title{Extreme intranight variability in the BL Lacertae object AO 0235+164}

\author{G.E. Romero \inst{1,}\thanks{Member
of CONICET}$^,$\thanks{Visiting Astronomer, Complejo Astron\'omico
El Leoncito, operated under agreement between
CONICET
and the National Universities of La Plata, C\'ordoba, and
San Juan.}, S.A. Cellone\inst{2, \star\star}, J.A. Combi\inst{1,
\star, \star\star}} \offprints{G.E. Romero}

\institute{Instituto
 Argentino de Radioastronom\'{\i}a, C.C.5,
(1894) Villa Elisa, Bs.\ As., Argentina \and Facultad de Ciencias
Astron\'omicas y Geof\'{\i}sicas, UNLP, Paseo del Bosque, 1900 La
Plata, Argentina}

\date{\today}
\titlerunning{Extreme variability in AO 0235+164}
\maketitle


\begin{abstract}
We present results of two-colour photometry with high time
resolution of the violently variable BL Lac object AO 0235+164. We
have found extreme intranight variability with amplitudes of $\sim
100$ \% over time scales of 24 hours. Changes of 0.5 magnitudes in
both $R$ and $V$ bands were measured within a single night, and
variations up to 1.2 magnitudes occurred from night to night. A
complete outburst with an amplitude $\sim30$ \% was observed
during one of the nights, while the spectrum remained unchanged.
This seems to support an origin based on a thin relativistic shock
propagating in such a way that it changes the viewing angle, as
recently suggested by Kraus et al. (1999) and Qian et al. (2000).


\end{abstract}

\section{Introduction}

The BL Lac object AO 0235+164 is one of the most intensively
studied blazars. It is a very compact source (e.g. Jones et al.
1984, Chu et al. 1996) which ejects superluminal components with
apparent velocities up to $\sim30c$ (see Fan et al. 1996 and
references therein). The object presents emission lines at a
redshift of $z=0.94$ and foreground absorption features at
$z=0.85$ and $z=0.524$, which have led several authors to study a
gravitational microlensing scenario for this source (e.g. Stickel
et al. 1988, Abraham et al. 1993).

The historical lightcurves at different optical frequencies of AO
0235+164 have been recently compiled by Fan \& Lin (2000). This
source is one of the most optically variable BL Lac objects (e.g.
Webb et al. 1988). Very rapid changes of its flux density have
been reported across the entire electromagnetic spectrum.
Quirrenbach et al. (1992), Romero et al. (1997), and Kraus et al.
(1999) have detected intraday radio variability. At optical
wavelengths, Rabbette et al. (1996), Heidt \& Wagner (1996), and
Noble \& Miller (1996) have found rapid changes of a few tenths of
a magnitude within a single night. Schramm et al. (1994) reported
extreme optical microvariability (1.6 mag in 48 hours), although
not with high time resolution. At high energies, AO 0235+164 has
also displayed significant variability (see Hartman et al. 1999).

In this Letter we present the results of two-colour optical CCD
photometry with high time resolution of AO 0235+164 carried out
with the 2.15-m CASLEO telescope. Our observations revealed one of
the most extreme variability events ever observed at optical
frequencies in any blazar. The observational results can be used
to shed some light on the mechanisms that generate the short-term
flux changes in this puzzling object.

\section{Observations and data analysis}

The observations were carried out during 6 consecutive nights in
November 1999 with the 2.15-m CASLEO telescope at El Leoncito, San
Juan, Argentina, as part of an extensive program of optical
monitoring of gamma-ray blazars. A cryogenically-cooled CCD camera
with an excellent cosmetics Tek-1024 chip (with a read-out-noise
of 9.6 electrons and a gain of 1.98 electrons adu$^{-1}$) was
used. Originally, several blazars were scheduled for observation
with a Johnson $V$ filter, but when it became obvious that A0
0235+164 was undergoing extraordinary brightness variations we
focused only on this source and incorporated observations with a
Kron-Cousins $R$ filter in order to achieve two-colour photometry
and a high time resolution. These conditions were obtained from
the third to the last night. Typical integration times were
$\sim100$ s. The CCD frames were bias subtracted and flat-fielded
using dome-flats and twilight frames in order to correct the
pixel-to-pixel variations of the CCD. Standard stars selected from
Landolt (1992) were also observed each night for magnitude
calibration.

Each target frame contained, in addition to the BL Lac object,
several stars which were used for comparison and control purposes
following the procedures described in detail in Romero et al.
(1999). Data reduction was made with the IRAF software package.
The aperture routine APPHOT was applied to perform the
differential photometry, and lightcurves were calculated as target
minus comparison star, as usual in this kind of studies (e.g.
Carini et al. 1991). In order to avoid possible spurious
variability introduced by seeing fluctuations that could affect
the blazar but not the point-like stars, we have followed the
recommendations of Cellone et al. (2000) and selected an 8-pixel
(6.5 arcsec) radius aperture. Checks using plots of the PSF FHWM
vs. time confirmed that the aperture was correctly selected, and
that the magnitude changes were real and not originated in varying
light contamination from the weak foreground galaxies induced by
small changes in seeing conditions.

The standard deviation ($\sigma$) of comparison minus control
stars was adopted as a measurement of the observational errors.
The average variability errors were $\sigma_{R}\approx0.011$ mag
and $\sigma_{V}\approx0.012$ mag, whereas the individual
photometric errors were typically $\sim0.004$ mag and $\sim0.005$
mag for the $R$ and $V$ bands, respectively.

Absolute calibration to the standard system was performed for AO
0235+164 and the field stars, tying all nights to the best
(photometric) one. The expected zero-point error is $0.01$ mag for
both $R$ and $V$ bands, although internal consistency in the $V-R$
colours should be better than this. The blazar's standard
magnitudes were corrected for Galactic extinction adopting
$E_{B-V}=0.08$ (Schlegel et al. 1998), and absolute flux densities
in both bands were calculated using the calibrations of Bessell
(1979). Tables with the resulting values of the flux densities of
AO 0235+164 from November 3 to November 8 1999 are available upon
request from G.E. Romero\footnote{romero@irma.iar.unlp.edu.ar}.

Spectral indices for the blazar were computed fitting a power-law
spectrum $F\propto\nu^{\alpha}$ and errors adequately propagated
from the flux values. Spectral index variability during the
observation span was additionally controlled using the scatter of
the spectral indices obtained for the field stars
$|\sigma_{\alpha}|\sim0.074$.

\section{Results}

The complete differential $V$-band lightcurve of AO 0235+ 164 is
presented in Figure 1 along with the corresponding curve for the
stellar comparison. The confidence of the variability is at a 26.3
$\sigma$ level. For the $R$ band the confidence is even higher:
30.9 $\sigma$. In Table 1 we present a summary of the results
obtained from the observations; from left to right we list the
observing band, the UT date, the variability error $\sigma$
obtained from the scatter of the field stars, the timescale of the
variability defined as $t_{\rm v}=\Delta F(dF/dt)^{-1}$, the
confidence level $C$ of the observed variability estimated as the
ratio between the scatter in the BL Lac lightcurve and the
observational errors: $C=\sigma_{\rm Bl}/\sigma$, the variability
amplitudes defined as in Heidt \& Wagner (1996): $Y=100[(\triangle
F)^2-2\sigma^{2}]^{1/2}/<\!F\!>$ (where $<\!F\!>$ is the averaged
flux density), the fractional variability index defined as
$FV=\triangle F/F_{\rm min}$, and, finally, the average flux
density in mJy.

\begin{figure}
\resizebox{\hsize}{!}{\includegraphics{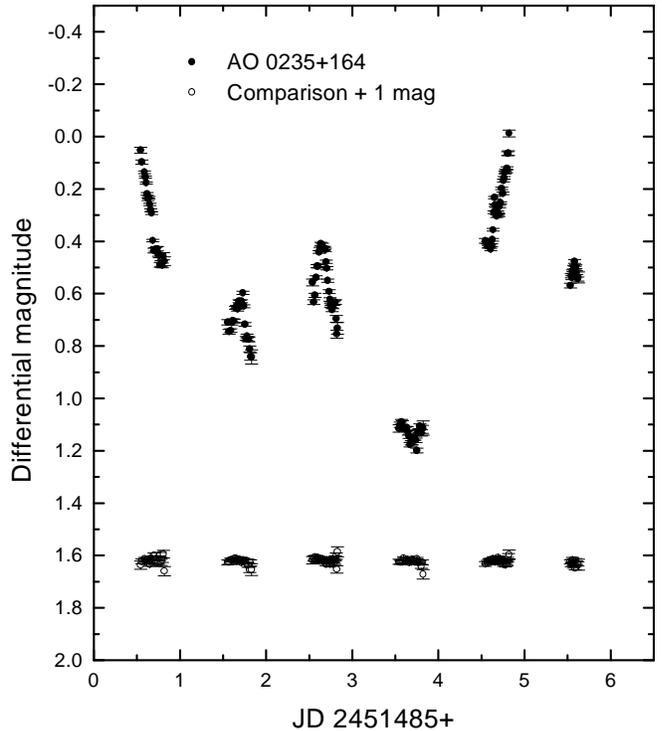}}
\caption{Differential lightcurve in the $V$ band for AO 0235+164
and comparison star during 6 consecutive nights in November,
1999.} \label{fig.1}
\end{figure}

The lightcurve shown in Figure \ref{fig.1} presents one of the
most extreme forms of optical variability reported in the
literature since the introduction of CCD cameras. Between the
fourth and the fifth nights the source brightness changed by $\sim
1.2$ mag in about 24 hours. Translated to flux density this means
a change of about 100\%: the source doubled its flux density in
one day. During the fifth night there was a variation of $\sim0.5$
mag in six hours. This extreme behaviour is even more violent than
the very strong outbursts observed in 1990 and 1991 by Noble \&
Miller (1996) in this same object, when intranight amplitudes of
$\sim0.25$ mag were registered.

\begin{table*}
\caption[]{Results of optical intranight variability observations
of AO 0235+164 \\ (entire campaign)}
\begin{tabular}{ l c c c c c c c }
\hline Band & UT Date  & $\sigma$  & t$_{\rm v}$ & $C$  &  $Y$ &
$FV$  & $<\!F\!>$  \\
     &          & (mag)     & (hours)     &      & (\%) &        &  mJy   \\
\hline
 $R$   & Nov 5-8, 1999 &  0.011 &  26.8    & 30.9 & 99.3  &  1.74 &  0.872 \\
 $V$   & Nov 3-8, 1999 &  0.012 &  26.8    & 26.2 & 109.4 &  2.04 &  0.563 \\
\hline
\end{tabular}
\end{table*}

In Figure 2 we show the spectral index evolution during the
observations (starting on the third night when the $R$ filter was
incorporated). The average index is $<\!\alpha\!>=-3.04$ (i.e.,
$<\!(V-R)_0\!>=0.68$), a steep value in agreement with the very
red colour indices quoted by V\'eron-Cetty \& V\'eron (2000). Some
day-to-day variability seems to be present in the spectral index,
but the confidence level ($C=1.65$) is too low as to be
conclusive. A Pearson's correlation analysis yields values of
$r=0.87$ for the linear correlation between both bands and the
spectral index, whereas it is $r\approx0.5$ for the stars, as
expected just from correlated observational errors. This could
mean a trend in the sense that the source becomes brighter when
the spectrum gets harder, as observed in other blazars (e.g.
Romero et al. 2000 and references therein). But, we emphasize, the
confidence level is too low as to allow any conclusion in this
respect.

\begin{figure}
\resizebox{6.5cm}{!}{\includegraphics{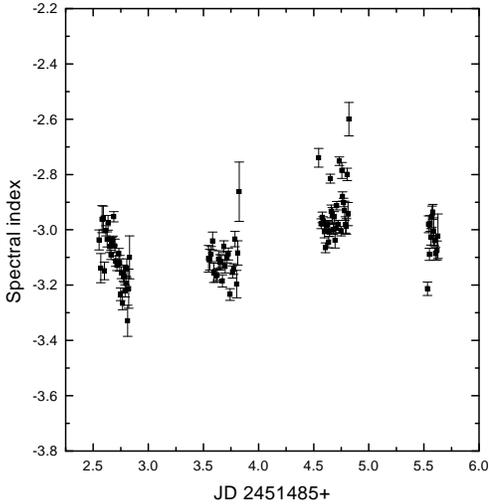}}
\caption{Spectral
index during the observations.} \label{fig.2}
\end{figure}

A correlation analysis using an interpolated correlation function
(ICF) gives a time lag of just 14.7 minutes for the entire
lightcurves, in the sense that variations appear first at higher
frequencies. However, the associated error, determined as in
Gaskell \& Peterson (1987), is 31.5 minutes, hence the variations can
be considered as simultaneous at both bands.

On the night of November 5, a complete outburst was observed. The
flux density in the $V$-band changed from $F_{\rm min}\approx0.46$
mJy to $F_{\rm max}\approx0.63$ mJy at a rate of $\langle
dF/dt\rangle\approx 0.04$ mJy/hr and returned to its original
value in 6 hours. This means a fluctuation with an amplitude
$Y\approx30.4$ \% in the $V$ band. In the $R$ band the variation
was similar: $Y\approx29.4$ \%, with identical timescale. The ICF
analysis is consistent with no time lag at all: $<\!{\rm
lag}\!>=1.3\pm5.7$ minutes. During the outburst, the spectral
index did not undergo significant changes; its average value was
$<\!\alpha\!>=-3.11$, with a scatter of $\sigma=0.09$ for the
entire night. In Figure 3 we show the lightcurves in both bands
for this event as well as the spectral index behaviour.

\begin{figure}
\resizebox{6.5cm}{!}{\includegraphics{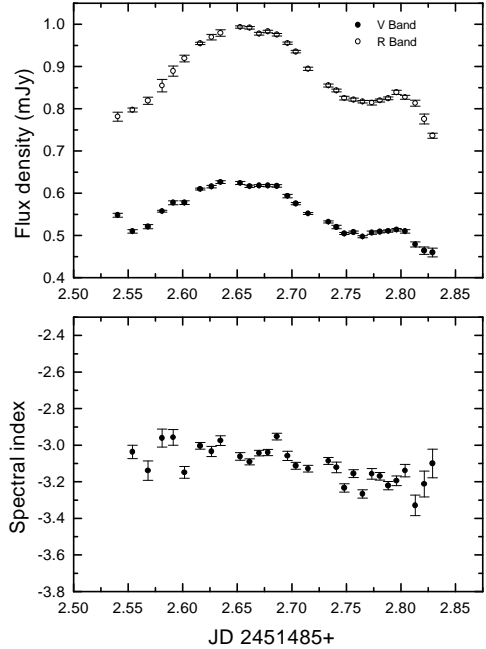}}
\caption{Outburst observed on November 5, 1999.} \label{fig.3}
\end{figure}


\section{Discussion}

Several models have been proposed to explain the rapid variability
of AO 0235+164. Interstellar scintillation is only relevant at
cm-wavelengths (e.g. Romero et al. 1997, Kraus et al. 1999), but
relativistic shocks and gravitational microlensing can produce
important changes in the flux density at optical wavelengths.
Accretion disk instabilities are also often invoked as a source of
optical variations in AGNs (e.g. Wiita 1996), but any accretion
disk model would have enormous difficulties to explain outbursts
of about 100 \% in a few hours. Extremely large and rapid optical
variations require the introduction of relativistic effects in
order to be explained (e.g. Marscher 1992). These effects are
present if an important part of the emission received by the
observer is originated in a superluminal feature, such as a thin
relativistic shock propagating down the jet of the source. Rapid
variability, then, arises when the shock interacts with density
inhomogeneities or turbulent features in the flow (e.g. Marscher
1990, Qian et al. 1991, Romero 1995). The shock evolution,
however, is not achromatic: radiative losses make the source
brighter when the spectrum flattens (e.g. Marscher 1997). This is
not the case for the rapid outburst shown in Figure \ref{fig.3},
although the evolution of the source during the entire
observational campaign could be consistent with this behaviour (as
we mentioned above the confidence on this point is not too high).
The rapid change observed on November 5 1999, on the contrary,
suggests a geometric origin because of its lack of spectral
variability and time symmetry.

Kraus et al. (1999) outlined a precessing jet model for AO
0235+164 where the trajectory of the shocked components is
determined by collimation in the magnetic field of the perturbed
beam, which can develop an helical configuration. This model has
been recently expanded by Qian et al. (2000) in order to include
aberration effects. If the relativistic source changes the viewing
angle due to its propagation through the helical path, the Doppler
factor should vary, introducing important changes in the observed
flux density. For a Lorentz factor $\Gamma=25$ (Qian et al. 2000),
a small change from $0\degr$ to $1\degr$ in the direction of the
shock with the line of sight would result in a change from 50 to
42 in the Doppler factor, and, consequently, in a flux variation
of $\sim70$ \% for the emitting component at all optical
wavebands. If the contribution from this component represents a
significant fraction of the total flux density, changes of $\sim
30$ \% can be easily obtained without variations in the spectral
index. If aberration effects are included, then the helical beam
model can also explain the radio behaviour observed by Kraus et
al. (1999), as it is shown by Qian et al. (2000), as well as
possible rotations of the polarization angle (e.g. K\"onigl \&
Choudhuri 1985, Romero et al. 1995a).

Alternatively to intrinsic models involving shocks, superluminal
gravitational microlensing (e.g. Romero et al. 1995b) could
explain the rapid and time symmetric outbursts observed in AO
0235+164. This is also an achromatic phenomenon; small lenses in
the interposed galaxies could amplify the emission from a thin
shock producing fast variations superposed to the longer ones,
which are due to the shock propagation. Although this mechanism
probably fails at radio frequencies because temperatures in excess
to the inverse Compton limit are demanded for the emitting region
in the blazar, at optical bands it cannot be ruled out at present
(see Rabbette et al. 1996 for additional details and numerical
estimates). The fact that a similar outburst was observed on the
night of November 4 (only in the V band) opens the question of how
frequently these outbursts occur. A large number could pose
problems to the microlensing interpretation because a high density
of small lenses would be required in the interposed galaxy,
leading to a "smearing out" of the individual outbursts.

Future simultaneous radio-optical observations with high time
resolution, as well as rapid polarimetric observations, would be
very helpful to improve our understanding of this extraordinary
blazar.

\begin{acknowledgements}

The authors acknowledge use of the CCD and data acquisition system
supported under US National Science Foundation grant AST-90-15827
to R.M. Rich. They are also very grateful to the CASLEO staff for
their kind assistance during the observations and to J.H. Fan and
S.J. Qian for useful remarks. This work has been supported by the
Argentine agencies CONICET (PIP 0430/98)
 and ANPCT (PICT 98 No. 03-04881), as well as by Fundaci\'on Antorchas
(through funds granted to GER and JAC).

\end{acknowledgements}


{}

\end{document}